\documentstyle[psfig,apjprepr,amssym,flushrt]{article}
\renewcommand{\plotone}[1]{\centerline{\psfig{figure=#1,width=9cm}}}
\renewcommand{\plottwo}[2]{\centerline{\psfig{figure=#1,width=7.7cm}
 \hskip 0.1 cm\psfig{figure=#2,width=7.7cm}}}

\newcommand{\citeauthor}[1]{\def\citeauthoryear##1##2##3{\rm ##1}\cite{#1}}
\newcommand{\citeyear}[1]{\def\citeauthoryear##1##2##3{\rm ##3}\cite{#1}}
\newcommand{\citeN}[1]{\citeauthor{#1} (\citeyear{#1})}
\newcommand{\citeNP}[1]{\citeauthor{#1} \citeyear{#1}}

\newcommand{\ujlisti}{
\itemsep=0 em
\parsep=0.5 em
\partopsep=0.25 em
\topsep=0 em}

\newcommand{\scri}{\scriptsize} 
\newcommand{\ptl}{\partial}
\newcommand{\dv}[2]{\frac{{\rm d}#1}{{\rm d}#2}}

\newcommand{\pdv}[2]{\frac{\partial #1}{\partial #2}}

\newcommand{\tfrac}[2]{{\textstyle\frac{#1}{#2}}}

\def\defdef{\buildrel \rm def \over =}
\def\uc{{u_c^{}}}
\def\B0{{B_0^{}}}\def\b0{{b_0^{}}}
\def\Be{{B_{\mbox{\scri e}}}}
\def\Bi{{B_{\mbox{\scri i}}}}
\def\nui{{\nu_{\mbox{\scri i}}}}
\def\re{{r_{\mbox{\scri e}}}}
\def\rt{{r_{\mbox{\scri CS}}}}
\def\rsp{r_0}
\newcommand{\rphi}{{r_0}}
\newcommand{\rphio}{{r_0}}
\def\alphanu{{\alpha_\nu^{}}}
\def\nablaad{{\nabla_{\mbox{\scri{ad}}}}}

\slugcomment{\it The Astrophysical Journal, \bf 485\rm , 398--408}
\lefthead{Petrovay \& Moreno-Insertis}
\righthead{Turbulent Erosion of Flux Tubes}

\begin{document}

\title{Turbulent Erosion of Magnetic Flux Tubes} 
\author{K. Petrovay and F. Moreno-Insertis}
\affil{Instituto de Astrof\'\i sica de Canarias, 38200 La Laguna (Tenerife), Spain}
\authoraddr{38200 La Laguna (Tenerife), Spain}
\authoremail{kpetro@ll.iac.es, fmi@ll.iac.es}

\begin{abstract}
Results from a numerical and analytical investigation of the solution of a
nonlinear axially symmetric diffusion equation for the magnetic field 
are presented for the case
when the nonlinear dependence of the diffusivity $\nu (B)$ on the magnetic
field satisfies basic physical requirements. We find that for sufficiently
strong nonlinearity (i.e.\ for sufficiently strong reduction of $\nu$ inside
the tube) a current sheet is spontaneously formed around the tube within one
diffusion timescale. This sheet propagates inwards with a velocity
inversely proportional to the ratio of the field strength just inside the 
current sheet
to the equipartition field strength $B_0/\Be$, so the lifetime of a tube
with constant internal flux density is increased approximately by a factor
not exceeding $B_0/\Be$, {\it even for infinitely effective
inhibition of turbulence inside the tube}. Among the applications of these
results we point out that toroidal flux tubes in the solar convective zone
are subject to significant flux loss owing to turbulent erosion on a
timescale of $\sim 1$ month, and that turbulent erosion may be responsible
for the formation of a current sheet around a sunspot. It is further
proposed that, despite the simplifying assumptions involved, our
solutions correctly reflect the essential features of the sunspot decay process.
\end{abstract}


\section{Introduction}
The relation of turbulence to magnetic structures is ambivalent. On the one
hand, any magnetic field concentration will tend to be
smeared out by the process of turbulent diffusion due to smaller scale
motions.  Since first proposed by \citeN{Sweet:turb.cond}, this process of
{\it turbulent magnetic diffusion} \/ characterized by a diffusivity
$\nu\sim lv$ (where $l$ is the scale of the turbulence and $v$ is its r.m.s.\
velocity) was and, despite repeated controversies, has remained a cornerstone
of mean field magnetohydrodynamics, observable e.g.\ in the form of a slow
dissolution of unipolar areas on the solar surface
(\citeNP{Sheeley+:first.combined}).  On the other hand, turbulent motions
tend to sweep the field lines together at convergence points of the flow
component perpendicular to them; thus, the field becomes concentrated in flux
tubes (\citeNP{Weiss:first,Proctor+Weiss:review}). This process of {\it flux
concentration} is thought to be aided by the fact that magnetic fields tend
to inhibit turbulent motions, and thus turbulent diffusion. 

In homogeneous stationary MHD turbulence, the fine structure of the magnetic
field is essentially determined by a dynamical equilibrium of these two main
processes. This leads to an intermittent field structure where at any given
instant of time most of the magnetic flux is concentrated into thin flux
tubes of about equipartition strength, which however possess a lifetime not
exceeding by much the turbulent correlation time. In inhomogeneous and/or
non-stationary situations, however, it may often happen that a flux tube gets
into an environment where its flux and/or field strength far exceed the
values corresponding to the local turbulent equilibrium described above. In
such situations one may expect that turbulent diffusion will dominate over
concentration, leading to the slow dissolution of the tube. 
A case in point are the sunspots, which are good examples
of such oversized and over-intense flux tubes (as compared to the local
turbulence scales). They are seen to decay in a few weeks after their
formation in a way very much reminiscent of turbulent diffusion
(\citeNP{Zwaan:ARAA}). A similar process of dissolution might be expected in
the case of the toroidal flux bundles lying near the bottom of the solar
convective zone: they may provide another example of oversized and
over-intense flux tubes if their sizes and strengths are inferred correctly
from flux loop emergence models.  A detailed analysis of such a turbulent
decay of magnetic flux tubes is however complicated by the fact that magnetic
fields with energy densities comparable to or exceeding the turbulent kinetic
energy density inhibit turbulence, and thereby reduce $\nu$ by a significant
amount. The exact form of the relation $\nu (B)$ is poorly known. 
(See \citeNP{Kichat+:quenching} for a formula valid under certain 
assumptions.) However,
from energy considerations it is clear that in the limit $B\rightarrow 0$ it
should tend to the kinematic value $\nu_0\sim lv$, while, in the strong field
limit $B\rightarrow \infty$, it must go to zero. At the equipartition field
value one may expect a reduction by a factor of order unity.

The strong inhibition of turbulence inside tubes with field strengths well
above the equipartition value (as in the case of the two examples mentioned
above) suggests that
%as inside the tubes the diffusivity is very small, 
the decay mostly advances by the ``gnawing'' action of the external turbulent
motions on the edges of the tube. This was first proposed by
\citeN{Simon+Leighton:supgr.obs}, who called this process the
{\it erosion} of the flux tube. They proposed that erosion by supergranular
motions was responsible for the decay of sunspots. Smaller-scale granular
motions may however be better suited for this task than the large-scale
supergranules.

This paper is an attempt at the detailed study of the turbulent erosion
process of magnetic flux tubes outlined above. For simplicity we consider an
axially symmetric magnetic flux tube and disregard the variation of physical
quantities along the axis of the tube, assuming that their lengthscale is
large compared with the tube radius.  For the nonlinear dependence of the
magnetic diffusivity $\nu$ on the field strength $B$ we consider a simple
analytic function satisfying the basic physical requirements. We note that the
magnetic field in general leads to an anisotropic diffusivity, so that,
strictly speaking, our $\nu$ is just the $rr$-component of the diffusivity
tensor in cylindrical coordinates. With these assumptions we write down and
numerically solve the nonlinear diffusion equation governing the problem in
\S~\ref{sec:diffprob}. We show that a steep {\it diffusion front} is easily
formed at the boundary of the tube when the internal field strength is large
compared to the equipartition value. This front slowly advances into the
tube, thus removing magnetic flux from it that is diffused outward into the
surrounding weak field region. \S~\ref{sec:analytical} contains an analytical
study of the problem, including the calculation of form--invariant,
propagating solutions (\S~\ref{sec:plpar}) and estimates for the velocity of
the front (\S~\ref{sec:w}) and of its thickness (\S~\ref{sec:sheet}).  In
\S~\ref{sec:applications} we will consider some implications of our results
for toroidal flux tubes rising in the convection zone and sunspots.

The problem of diffusion with a diffusivity dependent on the variable being
transported appears also in other physical and astrophysical contexts such as
turbulence propagation or viscous gravity currents (\citeNP{Barenblatt};
\citeNP{Gratton+Minotti}). An interesting case
is heat conduction in a plasma with conductivity, $\kappa$, given by
Spitzer's formula ($\kappa \propto T^{5/2}$). The strong dependence on $T$
gives rise in this case to {\it conduction fronts}, which are relatively
abrupt transitions from high to low temperatures. In that case, the front
becomes steeper toward low temperatures. In the magnetic case, in contrast,
the diffusivity is {\it suppressed} by the higher fields. Hence, we expect
the {\it magnetic diffusion front} (i.e.\ the current sheet) to become 
steeper toward higher field
intensities, as in  a {\it mesa}-like mountain profile (cf.\ Fig.~1$d$ below). 
An important
difference between the thermal and magnetic problems is the possibility of a
stationary situation: in the thermal problem one can compensate for the
energy being conducted with adequate amounts of heating and cooling on the
hot and cold ends, so that a stationary situation may ensue (as found in
different astrophysical contexts). In the problem of a  magnetic diffusion
front around a flux tube, in contrast, there are no source terms for the
magnetic flux inside the tube, so stationarity can only be achieved, if at
all, through an external advection term which brings back to the tube the
magnetic flux that the ohmic diffusion is trying to remove from it.

\section{The nonlinear diffusion problem}\label{sec:diffprob}

\subsection{Formulation of the problem} \label{sec:formulation}
As a first approach to the problem of the nonlinear diffusion of magnetic
field from a magnetic flux tube, we set out to solve a nonlinear
one-dimensional diffusion equation with axial symmetry.  The
one--dimensionality may limit its application to actual solar problems (see
Sect.~\ref{sec:applications}), but it has the advantage of retaining the main
nonlinear effects while keeping the problem relatively simple. The diffusion
equation for the magnetic field in cylindrical coordinates reads
\begin{equation}
\pdv{}t\left(rB\right)=\pdv{}r\left[r\;\nu(B)\;\pdv{B}r\right]
\; ,  \label{eq:diff}  \end{equation}
where the notations are $r$ for radius, $t$ for time, $B$ for magnetic flux
density and $\nu$ for magnetic diffusivity.
The inhibiting action of the magnetic field on turbulence is contained in the
form of the function $\nu(B)$ which should therefore have the following
properties:
\begin{description}
\item[(a)] $\lim^{}_{B\rightarrow 0}\nu =\nu_0$ where $\nu_0$ is the 
unperturbed value of the diffusivity
\item[(b)] $\nu/\nu_0\in{\cal O}(1)$ and $(1-\nu/\nu_0)\in{\cal O}(1)$ if $B\sim
\Be$, where $\Be$ is that value of the field intensity for which $\nu$ is
reduced by 50\%. We can  
expect $\Be$ to be order of the equipartition flux
density corresponding to a magnetic energy density that equals the kinetic
energy density of turbulence.
\item[(c)] $\nu/\nu_0\ll 1$ for $B/\Be\gg 1$ (assuming that the
molecular diffusivity is negligible)
\end{description}
A simple function satisfying the above requirements is
\begin{equation} \nu(B)=\frac{\nu_0}{1+|B/\Be|^{\alphanu}_{}} 
   \label{eq:nuexpr} 
\end{equation}

\noindent The parameter $\alphanu$ determines the steepness of the
diffusivity cutoff 
near $\Be$. All the results presented below were computed with
the above form of $\nu(B)$; however, test runs with other forms of $\nu(B)$
compatible with conditions (a) through (c) yield qualitatively similar
findings. Furthermore, in \S~\ref{sec:sheet} below we 
will argue that the general behavior of the solutions should be similar for
all forms of $\nu(B)$ obeying the criteria listed above.

As initial condition we again choose simple analytic $B(r)$ profiles of the
form
\begin{equation} B(r)=\frac{B_0}{1+(r/r_0)^{\alpha_B}}  \label{eq:B0expr1} \end{equation}
with $\alpha_B=22$ (to model a tube with constant field strength inside), or
\begin{equation} B(r)=B_0\exp (-r^2/r_0^2)  \label{eq:B0expr2} \end{equation}  
for a tube with a field profile gradually decreasing outwards.

\subsection{Numerical solutions} \label{sec:numsol}
Equation (\ref{eq:diff}) is a nonlinear flux-conserving equation that may be solved by
standard finite-differencing techniques. A two-step Lax-Wendroff scheme was
applied for this purpose. In the numerical calculations, and whenever 
dimensionless quantities are implied in the paper, the following units were
used:  $r_0=\nu_0=\Be=1$. (Note that the unit of time is thus the diffusive
timescale $r_0^2/\nu_0$.)

For the solution one should also specify the boundary conditions. At $r=0$
the boundary condition is set by the requirement of symmetry, while at the
outer boundary we experimented with different forms of the boundary condition
to find that if this boundary is sufficiently far away (at $r=10$) the form
of the boundary condition exerts a negligible influence on the results in the
physically interesting regime $r<2.5$ (in the figures, only this interior
regime is shown). Consequently, for all calculations we set the outer
boundary at $r=10$ as $\ptl B/\ptl r=0$. The outer fictive points necessary
for the Lax-Wendroff method were computed with the neglect of third
derivatives; this method preserves the second-order accuracy of the scheme.

%\placefigure{fig1}
\begin{figure}[ht]
\plottwo{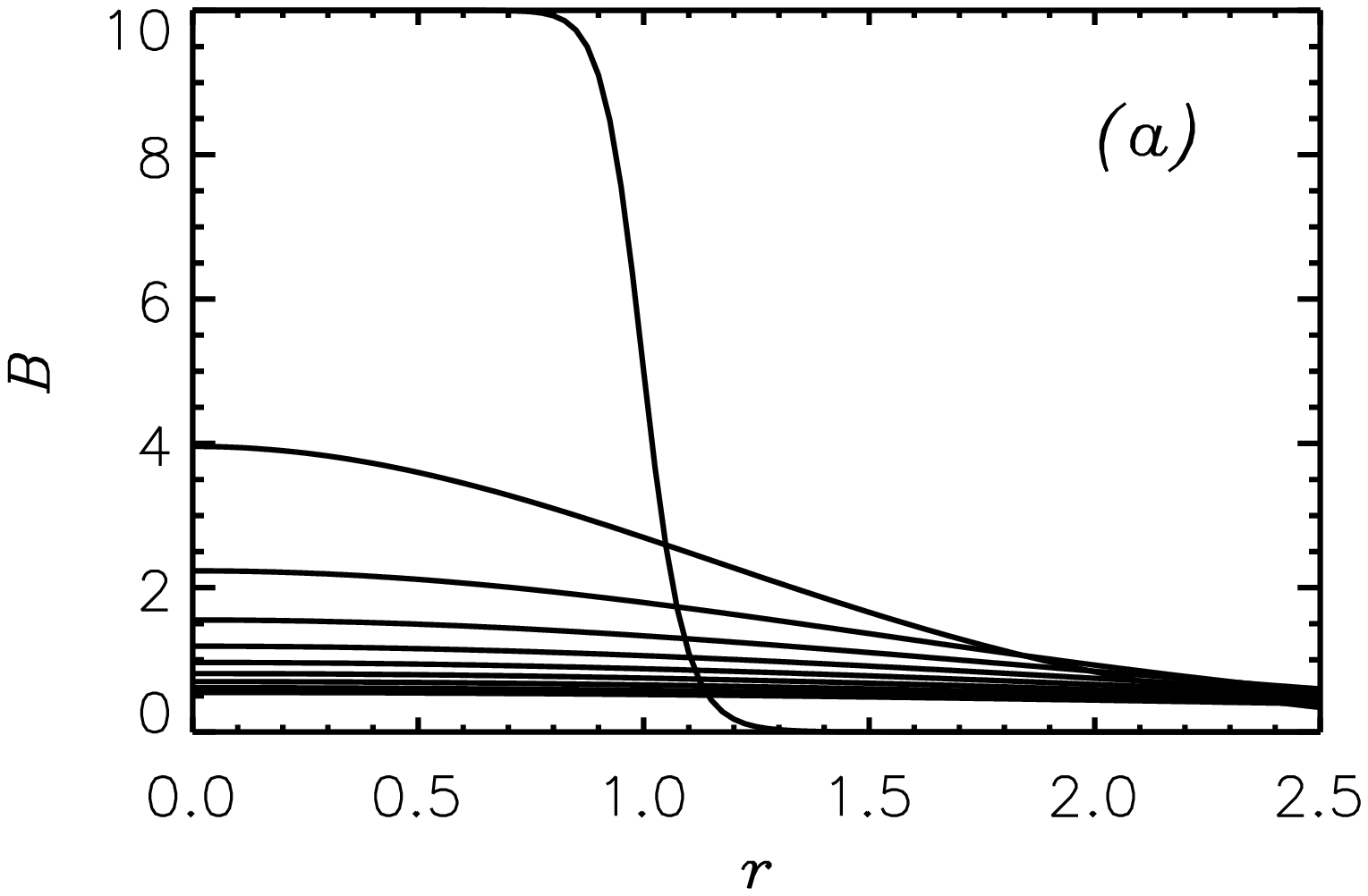}{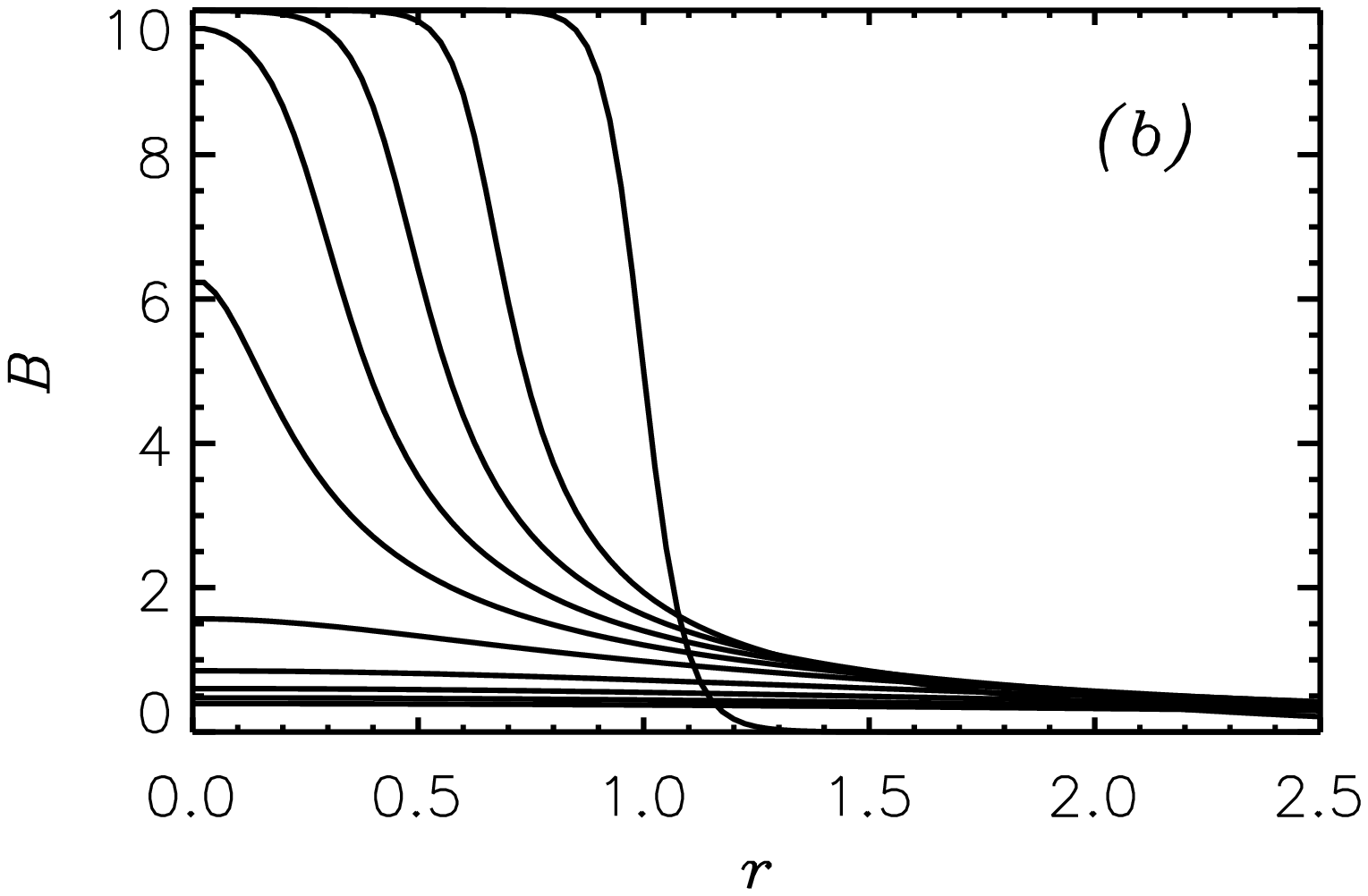}
\plottwo{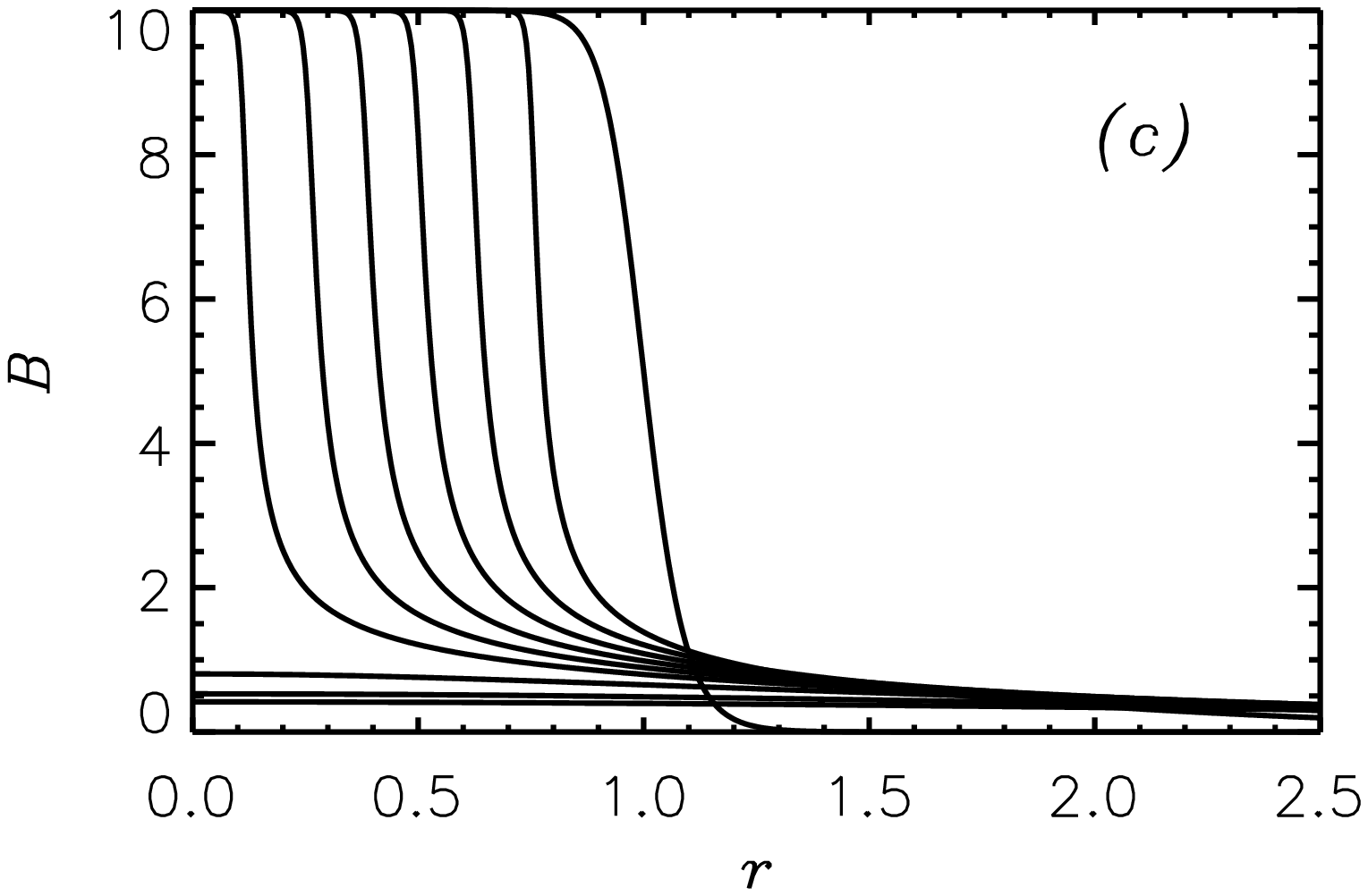}{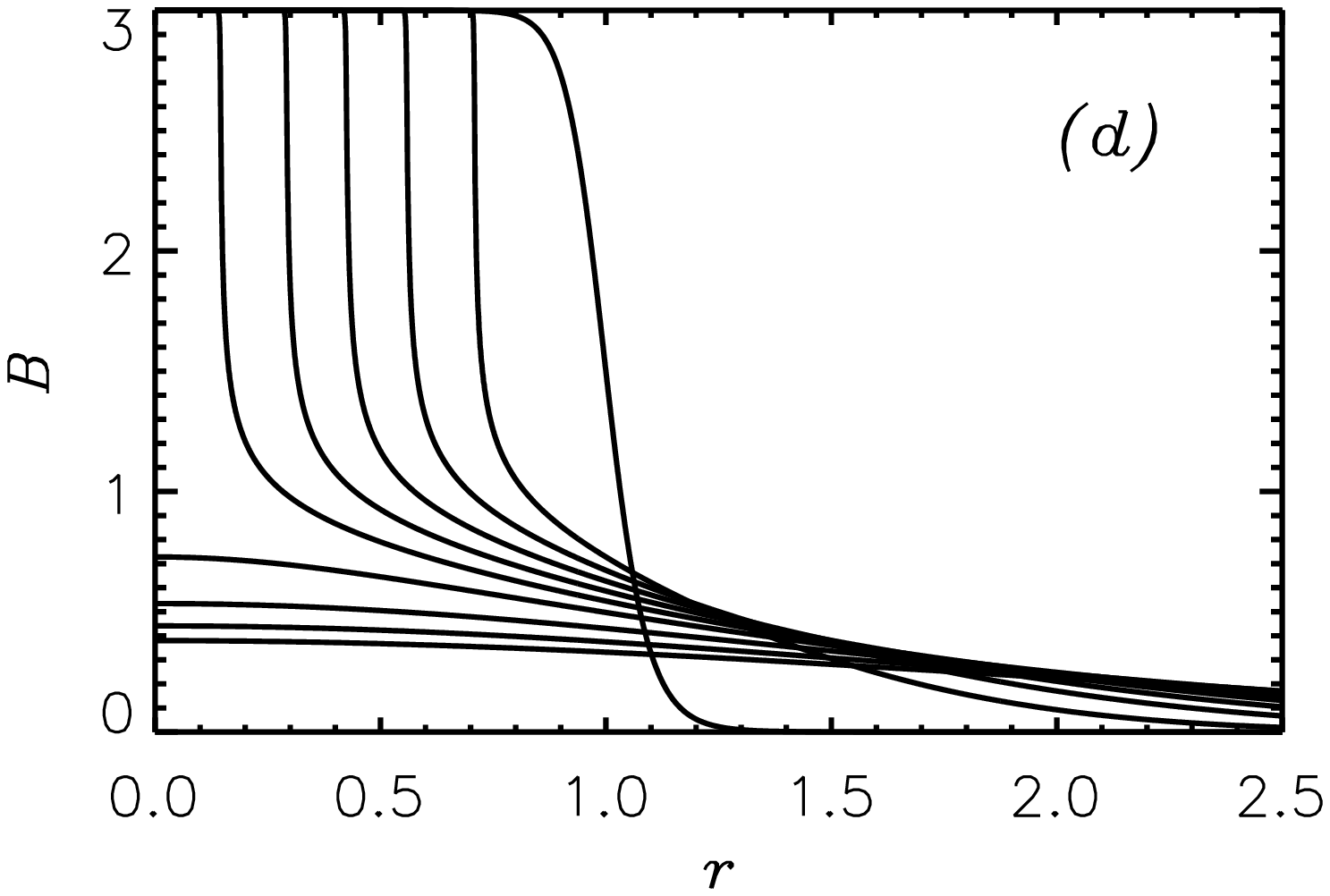}
\caption{Nonlinear turbulent diffusion of a magnetic flux tube. 
Shown are snapshots of the magnetic field profiles 
at 10 equidistant instants of time (with timesteps $\Delta t_0$, in units of 
the diffusive timescale)
for a magnetic diffusivity of the form (2) 
%(\ref{eq:nuexpr}) 
with $\alphanu = 0$, $\Delta t_0=0.5$ (linear limit,
panel \it a\/\rm ), $\alphanu = 2$, $\Delta t_0=1$ \it(b), \rm $\alphanu = 3$, 
$\Delta t_0=1$ \it (c), \rm and $\alphanu = 7$, $\Delta t_0=0.25$ \it (d). \rm 
Nondimensional units as defined in Section 2.2   
\label{fig1}}
\end{figure}

The stability criterion $\Delta t\le(\Delta x)^2/2\nu$ seriously limits the
timestep; therefore, owing to the finite available CPU time, the maximal
spatial resolution that could be attained was $8000$ grid points (uniformly
distributed in the range $0\le r\le10$), with a corresponding timestep of
$10^{-5}$.

\begin{figure}[ht]
%\plottwo{fig1a.eps}{fig1b.eps}
%\plotfiddle{fig2.eps}{10 cm}{0}{100}{100}{0}{0}
\plotone{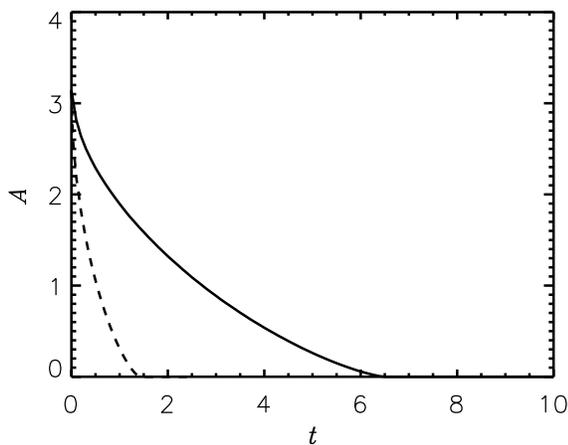}
\caption{
The area enclosed within the current sheet as a function of time. 
The cases $\alphanu=3$ (solid) and $\alphanu=7$ (dashed) are shown 
(corresponding to panels $c$ and $d$ in Fig.~1).  %\ref{fig1}
\label{fig1area} }
\end{figure}

Figure \ref{fig1} presents the time evolution of the field profiles with
initial condition given by (\ref{eq:B0expr1}) with $\alpha_B^{} = 22$ and for
different values of the coefficient $\alphanu$ in equation~(\ref{eq:nuexpr}),
namely $\alphanu=0$ (fully linear case, panel $a$) and other three which
yield increasingly steep profiles (panels $b$, $c$ and $d$, with $\alphanu =
2, 3, 7$, respectively).  For low $\alphanu$ the solution is still similar to
the linear constant-diffusivity case, viz., it is characterized by a
monotonic decrease of field gradients, smoothing out the field distribution,
and by a gradual decrease of the flux density at $r=0$. However, for higher
values of $\alphanu$ a qualitatively new behaviour sets in: in a short
interval of radius the field gradient (or equivalently the current density
$j_\phi = -4\pi/c\; \ptl B/\ptl r$) greatly increases, i.e. a {\it current
sheet 
\/} is formed spontaneously. The flux density on the axis remains constant
until the arrival of the current sheet and then it suddenly drops to nearly
zero. The sheet moves inwards with an essentially constant speed, thereby
leading to a nearly parabolic decay law for the flux and area inside it
(cf.~Fig.~\ref{fig1area}). Within about the equipartition radius, 
$r \lesssim \re$,
where $B(\re)=\Be$, the overall form of $B(r,t)$ is consistent with a
dependence of the form $B(u)$ where $u$ is the distance from the
current sheet, with no explicit time dependence. 
In what follows we will call solutions of this type {\it erosional
solutions}, \rm a name which reflects the fact that the formation and inward
propagation of the current sheet is due to the eroding action of turbulent
magnetic diffusivity on the outskirts of the tube, while in the tube interior
diffusion is practically absent.

%\placefigure{fig1area}

The thickness of the current sheet quickly decreases with increasing values
of $\alphanu$. This limits the possibility of calculating numerically the
solutions for 
high $\alphanu$. The highest value for which we still had good resolution,
$\alphanu=7$, was reached by keeping $B_0$ at a moderate value of 3 
(Fig.~\ref{fig1}$d$). In this case, the current sheet extends just to
$B \sim \Be$; from there outward there is a range of radii where
the solution quickly approaches the solution of a linear diffusion
problem. For higher values of $\alphanu$, this transition range shrinks to
basically just the point where $B=\Be$. In that case a near discontinuity in
the derivative of the field profile (i.e., in the current density) appears. 

\begin{figure}[htbp]
%\plottwo{fig1a.eps}{fig1b.eps}
%\plotfiddle{fig2.eps}{10 cm}{0}{100}{100}{0}{0}
\plotone{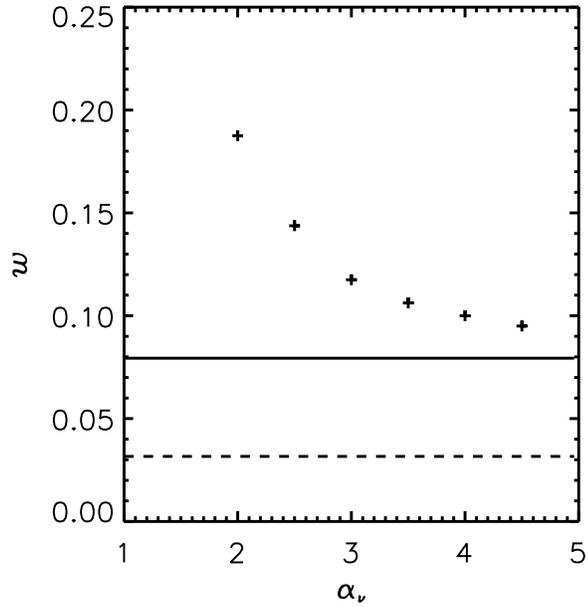}
\caption{
Current sheet velocity $w$ as a function of $\alphanu$
for models with $B_0=10$, $\alpha_B=22$ with the limits given by
 eqs.\ (16) %\ref{eq:wlimit2}
 [solid] and (15) %\ref({eq:wlimit1})
 without the $3^{-1/2}$ factor [dashed]
 \label{fig2} }
\end{figure}

\begin{figure}[htbp]
%\plottwo{fig1a.eps}{fig1b.eps}
%\plotfiddle{fig2.eps}{10 cm}{0}{100}{100}{0}{0}
\plotone{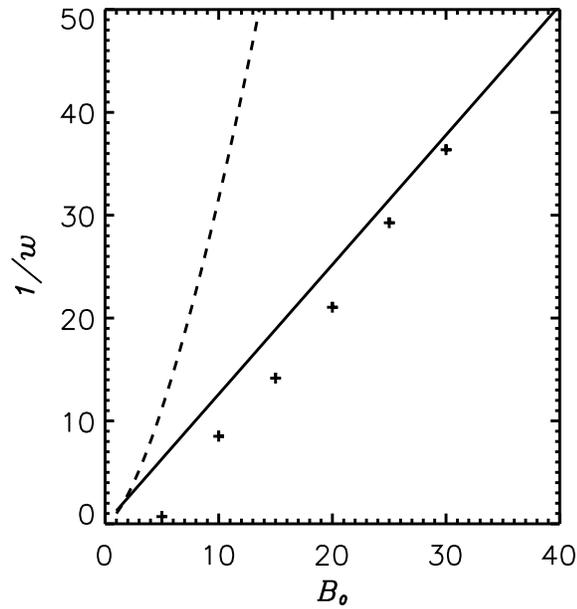}
\caption{
Inverse of current sheet velocity $1/w$ as a function of
$B_0$ for models with $\alphanu=3$, $\alpha_B=22$, with the limits given by
 eqs.\ (16) %\ref{eq:wlimit2})
 [solid] and (15) %\ref{eq:wlimit1}) 
 without the $3^{-1/2}$ factor [dashed]. 
\label{fig3} }
\end{figure}

Figures \ref{fig2} and \ref{fig3} show the propagation velocity $w$ of the
current sheet as a function of $\alphanu$ and $\B0$, respectively. The
velocity $w$ was determined by measuring the horizontal separation of
consecutive $B(r)$ curves at different times and $B$ values inside the
current sheet.  For a constant internal field strength, $1/w$ is a good
estimate of the total lifetime of the tube. While the lifetime increases
approximately linearly with $B_0$, it seems to saturate to a finite
asymptotic value with increasing $\alphanu$ when $B_0$ is constant.  This
asymptotic value is approximately $B_0$. 

%\placefigure{fig2}

%\placefigure{fig3}

\begin{figure}[htbp]
%\plottwo{fig1a.eps}{fig1b.eps}
%\plotfiddle{fig2.eps}{10 cm}{0}{100}{100}{0}{0}
\plotone{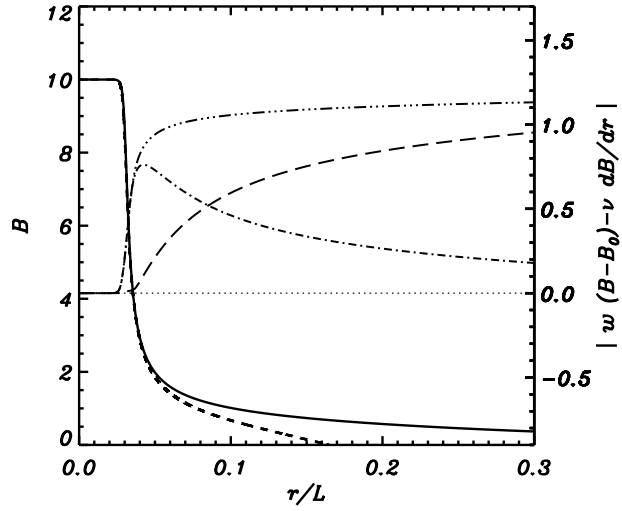}
\caption{
\label{fig1mff}
Plot of $ w (B-\B0) - \nu(B) \partial
B/\partial r$ (long-dashed curve) for an intermediate time of the solutions
of Fig.~1 with $\alphanu=3$. In the figure, the numerical solution for $B$ (solid line) and the
corresponding analytical propagating solution (short dashes) are shown with
ordinates on the left y-axis. Two further curves are shown with ordinates on
the right y-axis: $- \nu(B) \partial B/\partial r$ (dash-dot) and $-w
(B-\B0)$ (dash-triple-dot). The constancy of the long-dashed curve to the
left of and within the current sheet is a direct consequence of the
form-invariance of the erosional solutions.}
\end{figure}

The magnetic flux that is being removed from the strong-field region is
diffused away outward from the current sheet. The rate of removal of flux is
roughly $2\pi\rt w \B0$, where $\rt$ is the radius of the
current sheet. Given the form invariance of the solutions shown in the panels
of Figure~\ref{fig1} with higher $\alphanu$'s, 
we can expect the flow of magnetic
flux across the current sheet to be essentially constant: flux cannot
accumulate (nor the contrary) within the sheet if it is to keep a constant
shape in time. Hence, the following quantity, the magnetic flux removal rate,
$[w B - \nu(B) \partial B/\partial r]$ should be constant in the whole
strong-field range and across the current sheet. This is borne out by the
numerical solutions: Figure~\ref{fig1mff} shows that the profile of that
quantity (long-dashed curve) is basically constant from $r=0$ out to the
neighborhood of $B=\Be$. 
Thereafter it
smoothly increases: in the diffusive part there cannot be a form invariance,
since the magnetic flux must increase along time to accommodate what is being
removed from inside.

%\placefigure{fig1mff}

\begin{figure}[htbp]
%\plottwo{fig1a.eps}{fig1b.eps}
%\plotfiddle{fig2.eps}{10 cm}{0}{100}{100}{0}{0}
\plotone{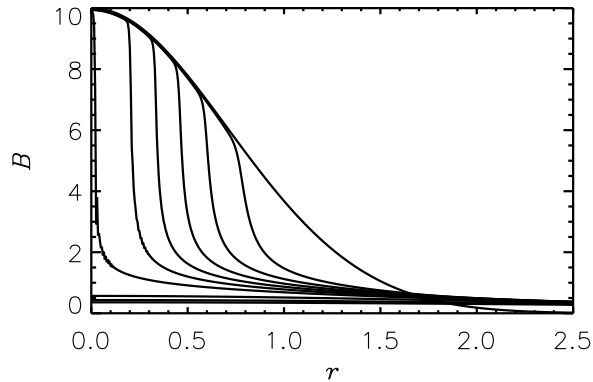}
\caption{
\label{fig1gauss}
Same as in Fig.~1
%\ref{fig1} 
but with a Gaussian profile as initial condition
[eq.~(4)], %\ref{eq:B0expr2}
$\alphanu=3.5$, $\Delta t_0=1$. 
The velocity of erosion is no longer constant, but,
rather, decreases for increasing field on the {\it strong end} of the
current sheet.}
\end{figure}

The existence of this class of erosional solutions is not limited to the case
of the initial conditions given by equation (\ref{eq:B0expr1}). Figure
\ref{fig1gauss} shows that such solutions are also found with sufficiently
high values of $\alphanu$ for other initial conditions, the only difference
being that for an internal magnetic field depending on $r$ the propagation
velocity of the current sheet is not constant anymore.

%\placefigure{fig1gauss}

%\placefigure{fig1kolmogorov}
\begin{figure}[htbp]
%\plottwo{fig1a.eps}{fig1b.eps}
%\plotfiddle{fig2.eps}{10 cm}{0}{100}{100}{0}{0}
\plotone{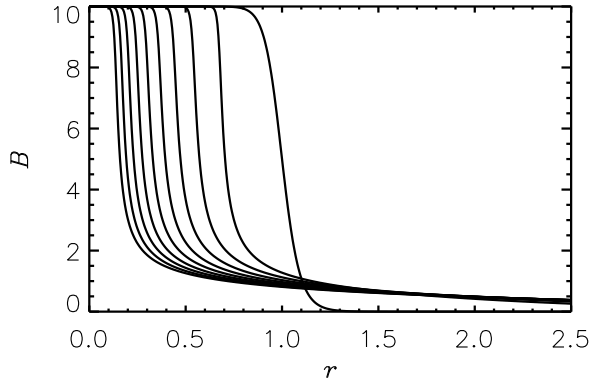}
\caption{
Same as in Fig.~1$c$
%\ref{fig1}
but using a cutoff for the size of the turbulent elements equal to the tube
radius and a Kolmogorov spectrum for the turbulence, with $\Delta t_0=2$. 
The value of $\nu_0$ in equation~(2)
%\ref{}
then becomes a function of the tube radius and the decay of the tube is 
strongly decelerated\label{fig1kolmogorov}}
\end{figure}

Up to this point we tacitly assumed that the value $\nu_0$ of the diffusivity
outside the tube is constant. This corresponds to a case where the size of
turbulent eddies is small compared to the size of the tube. In the
alternative case when the turbulent eddies are large, only the small-scale
turbulence (with scales below the tube radius) may contribute to the
diffusivity, while larger-scale motions will simply lead to the translational
motion of the tube as a whole. Assuming a Kolmogorovian spectrum, the
diffusivity scales as $\rt^{4/3}$. 
A solution with such a {\it rescaling} of $\nu_0$ in each timestep to
the appropriate value of $\rt$ (assuming $\nu_0=1$ at $t=0$) is illustrated
in Figure \ref{fig1kolmogorov}. The decay of the tube suffers a strong
deceleration 
with time, making the total lifetime of the tube effectively infinite;
however, a significant fraction (e.g.\ 50 or 90 percent) of the total flux is
still lost on a timescale comparable to that of the solution without
rescaling.

\section{Analytical solutions and estimates}\label{sec:analytical}
Among the features of the numerical solutions
discussed above, some merit particular attention. In the present subsection
we clarify and substantiate some of them using analytical
considerations. This allows us as well to extrapolate them to the numerically
unattainable limit $\alphanu\rightarrow\infty$.

\subsection{Propagating solutions} \label{sec:plpar} 
The numerical results show that in the case of a constant internal flux
density, for sufficiently strong nonlinearity (i.e.\ for sufficiently high
values of $\alphanu$ and/or $B_0$), the solution in the regime $B\ga\Be$
approximates a form-invariant {\it propagating solution} 
of type $B=B(r+wt)$, with $w$ approximately constant.  This invariant shape
is reached within a diffusion timescale $r_0^2/\nu_0$.

This prompts the question of the existence of purely propagating,
form-invariant solutions of the nonlinear diffusion equation
(\ref{eq:diff}). To obtain those, we go to the cartesian
limit of the equation (with spatial coordinate $x$) and try to obtain
solutions which depend on $t$ and $x$ through the combination $u \defdef x +
w t$, with $w$ a constant value (a positive $w$ means a solution moving
toward smaller spatial coordinate, as in the numerical solutions). The
corresponding equation is  
\begin{equation}\frac{d\ }{du}\left[-w B + \nu(B)\frac{d B}{du}\right] = 0 \;. 
\label{eq:propsoln} \end{equation}  
This equation has as immediate solution the following class of functions:
\begin{equation} u-\uc = \int \frac{\nu(B)}{K + w B} \;dB \;,
\label{eq:intforminv} \end{equation}
with $\uc$ and $K$ two integration constants. 
Equation~(\ref{eq:intforminv}) gives
$B$ implicitly as a function of $u-\uc$. This class of solutions
(\ref{eq:intforminv}) has a flat, exponential asymptote for $B \rightarrow
-K/w$, which, therefore, may serve to describe the flat region shown by the
numerical solutions of the previous section toward $r\rightarrow 0$. We then
write $-K/w = \B0$, introduce the notation $b=B/\Be$, $\b0 = B_0/\Be$, and use
the form (\ref{eq:nuexpr}) for $\nu$ so that the solution (\ref{eq:intforminv})
becomes
\begin{equation} u-\uc = \frac{\nu_0^{}}{w} \int \frac{db}{(1+|b|^{\alphanu})(1-b/\b0)} \;.
\label{eq:intforminvb} \end{equation}
\noindent The integral in equation~(\ref{eq:intforminvb}) can be easily carried
out analytically for integer $\alphanu$ (although the solution becomes
somewhat unmanageable for $\alphanu \gtrsim 5$) or, numerically, for
arbitrary $\alphanu$. For instance, for $\alphanu = 2$ one obtains
\begin{equation} u-\uc =  \frac{\nu_0}{w}\;\frac{\b0}{1+\b0^2}
\left[\b0\,\hbox{atan} b 
+ \log\frac{(1+b^2)^{1/2}}{1-b/\b0} \right] \;. \label{eq:analsol2} \end{equation}

%\placefigure{fig8}
\begin{figure}[htbp]
%\plottwo{fig1a.eps}{fig1b.eps}
%\plotfiddle{fig2.eps}{10 cm}{0}{100}{100}{0}{0}
\plotone{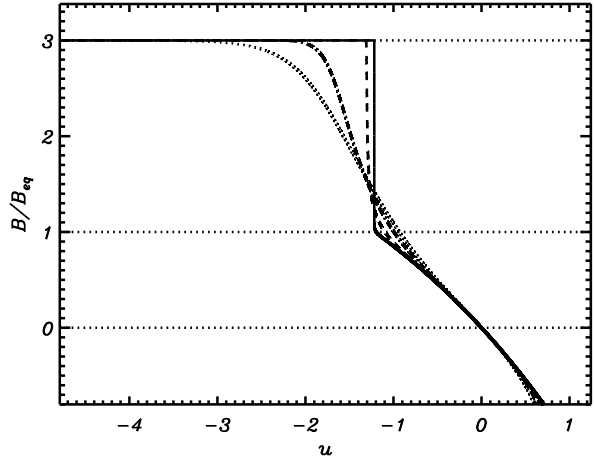}
\caption{
Analytic solutions for a plane parallel form-invariant
propagating solution of the nonlinear diffusion equation (6) with $B_0=3$ and
$\alphanu=2$ (dotted), $\alphanu=3$ (dot-dashed), $\alphanu=7$ (dashed) and
$\alphanu=50$ (solid). $u$ is dimensionless in this figure, unit is
$\nu_0/w$. The near discontinuity of the current density at the lower end of
the current sheet is clearly discernible in the steepest case. 
\label{fig8}}
\end{figure}

\noindent The form of the solutions (\ref{eq:intforminvb}) is illustrated in
Figure \ref{fig8} for four different cases, namely $\alphanu = 2$ (dotted),
$\alphanu = 3$ (dot-dashed), $\alphanu = 7$ (dashed) and a very steep case,
$\alphanu=50$ (solid), which illustrates well the limit
$\alphanu\rightarrow\infty$.  The increasing nonlinearity makes the
transition between the diffusive and non-diffusive regions increasingly
steep, similarly to the case of heat conduction with nonlinear conductivity
coefficient. For $\alphanu = 50$, in particular, the current sheet is very
narrow, with a sharp transition to the asymptotic horizontal regime. For $B
\lesssim \Be $, the four solutions are qualitatively similar and approach a
linear diffusive solution.  In these plane parallel solutions, there is no
natural lengthscale imposed by the equations: solutions exist for arbitrary
$w>0$ values, independently of $B_0$. A change in $w$ does not change the
shape of the solution, but only its lengthscale, which is $\nu_0/w$.  In
contrast, in the actual axisymmetric problem a length scale is set by the
radius $r_0$ of the initial tube and $w$ can no longer be specified
arbitrarily (cf.\ Section \ref{sec:w} below).

%\placefigure{fig9}
\begin{figure}[htbp]
%\plottwo{fig1a.eps}{fig1b.eps}
%\plotfiddle{fig2.eps}{10 cm}{0}{100}{100}{0}{0}
\plotone{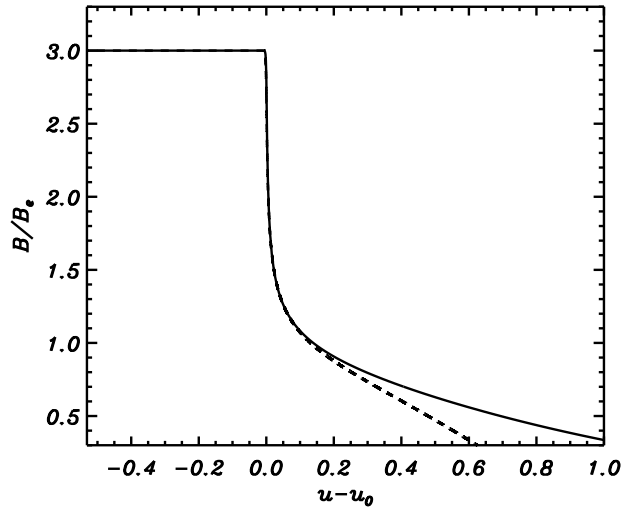}
\caption{
Comparison of the analytical (dashed) and numerical
(solid) solutions for $B_0=3$ and $\alphanu =7$. The analytical solution
was computed with the $w$ value determined from the numerical solution,
and the arbitrary horizontal offset was set to optimize the agreement.
\label{fig9}}
\end{figure}

We expect the propagating solutions to fit the numerical solutions of the
previous section increasingly well the higher the value of $\alphanu$  and
$\b0$ and,  at any rate, for $b\gtrsim 1$. 
The goodness of the fit can be seen in Figure~\ref{fig9}, which contains
the numerical (dotted)  and analytical (dashed) solutions for the case
$\alphanu=7$ with $B_0 = 3$. The fit is excellent in the whole non-linear 
region. 

The propagating solutions cannot approximate the numerical solution well
outside the current sheet for fundamental reasons: the current sheet is a
region where magnetic flux is being taken away from the tube and transported
outwards. To have a full form-invariant solution, there should be a
mechanism to exactly dispose of this flux in the outer regions of the
domain. The latter can be, for example, a flux-shedding boundary condition or
some internal device that causes the loss of magnetic flux (which should play
a role like the cooling in a thermal problem).  We have none of these, and,
as a consequence: (a) the numerical solution is not form--invariant in the
weak-field range and (b) the analytical solution goes to negative field
values at some distance of the flux tube (see Fig.~8). 

The establishment of a form-invariant propagating region along the time
evolution of the system may then be qualitatively understood as follows. The
high diffusivity in the outskirts ($B\la \Be)$ of the tube removes the flux
in that domain in about one diffusive timescale $r_0^2/\nu_0$, while the
strong internal field at $r<r_{CS}$ effectively inhibits any significant flux
loss from the internal region over such short timescales. As a consequence, a
steep field gradient (i.e.\ a current sheet) forms. The gradient increases
until the smoothing effect resulting from the diffusive flux loss can balance
the non-linear steepening. This is exactly the balance responsible for the
form-invariant nature of the exact analytic solution of equation
(\ref{eq:propsoln}), which is therefore a good approximation of the solution
in the regime $B\gtrsim \Be$.  This property of the erosional solution may be
utilized in deriving a formula for the thickness of the current sheet (cf.\
Section \ref{sec:sheet} below).

\subsection{Current sheet velocity}   \label{sec:w}
From the numerical solutions, the velocity $w$ of the inwards propagation of
the solution (and of the current sheet in particular) appears to increase
linearly with $B_0$ and to become asymptotically independent of $\alphanu$
for high $\alphanu$ values, i.e.\ for increasingly effective magnetic
inhibition of turbulence.  This implies that the lifetime of flux tubes
remains finite even in the limit of $\alphanu\rightarrow\infty$, i.e.\ for
infinitely effective inhibition of turbulence inside the tube!  Although
Fig.\ \ref{fig2} strongly suggests such a behaviour, a strict proof of 
this result clearly cannot be given on the basis of numerical results alone.
Below we demonstrate with approximate analytical methods 
that a lower
limit for $w$ indeed exists. Here we concentrate on the case of a flux tube
with constant internal flux density and a corresponding form-invariant
propagating $B(u)$ profile in the erosional solution. It is however clear
that for tubes with the same value of $B_0$ but outwards decreasing field
strength the efficiency of the erosion process can only increase.

The analytical results for the plane parallel case (cf.\ Section
\ref{sec:plpar} above) are of no help in the present problem since these
solutions exist for any value of $w$. It is clear from a dimensional analysis
that the most generic combination of the parameters entering the problem
($\nu_0$, $\alphanu$, $r_0$, $\Be$, $B_0$) with velocity dimensions is
$f(B_0/\Be, \alphanu)\nu_0/r_0$. %, which in turn involves the tube radius. 
Our task consists of determining the form of the function
$f(B_0/\Be,\alphanu)$ in the limit $\alphanu\rightarrow\infty$.

To obtain that function, note that in that limit the current sheet becomes
infinitely narrow, is placed precisely at the equipartition radius, $r=\re$,
and coincides in its full 
extent with the current sheet of the corresponding form-invariant propagating
solution.   
Moreover, the transition
between the current sheet and the $\nu=\nu_0$ regime occurs sharply at
$B = \Be$. In fact, $\nu(B)$ will have a near discontinuity at
$B=\Be$ and so will the derivative $\partial B/\partial r$. 
We saw in the previous section that the form invariance implied  
that the flow of magnetic flux, $w B - \nu(B)\frac{d B}{dr}$, was constant in
the whole form-invariant section of the solution down to the lower end of the
current sheet (and slowly departs from the constant value when going outward
from it). In our case, this means that that quantity is constant down to
$B=\Be$.  
Using all these
facts, we can now obtain a first estimate for the velocity of the current
sheet. Equating the values of that quantity on both ends of the current
sheet, we obtain
\begin{equation}
w\,\B0 = w \Be - \nu_0\,\left(\pdv{B}r\right)^{}_{B=\Be}
\;.\label{eq:fluxcont}  \end{equation}

\noindent where the last
derivative is meant as the limit value when one approaches $B=\Be$ from
{\it outside} the current sheet. This leads to 
\begin{equation} w=\frac{\nu_0}{r_1}\frac{\Be}{\B0-\Be} ,  \label{eq:wr1}    \end{equation}
with
\begin{equation} \frac 1{r_1} \defdef -\frac 1{\Be}
   \left(\pdv Br\right)^{}_{B=\Be} \;. \label{eq:r1def}
\end{equation}

\noindent The question is now how to determine $r_1$ in equation~(\ref{eq:wr1}).  
An upper limit can be set by noting that outside the current sheet,
i.e.\ for $r>\re$, $\ptl^2B(r)/\ptl r^2>0$, so that the straight line
$B=\Be\left[1-(r-\re)/r_1\right]$, tangential to 
the solution at $\re$, certainly runs below the actual field distribution. 
The latter, then, contains more magnetic flux than the straight
line. Using the approximation $\Phi_0\simeq\pi r_0^2 B_0$ where 
$\Phi_0$ is
the total magnetic flux in the whole domain, we have
\begin{equation} 2\pi \Be\int^{\re+r_1}_{\re}\left[1-(r-\re)/r_1\right]r\,dr 
   < \pi B_0(r_0^2-\re^2)\;.
\label{eq:fluxint}
\end{equation}
This inequality should be valid at any stage after the moment when the
form-invariant propagating solution finally sets in and $r_1$ settles to
a basically constant value.  In order to determine the
constraint on $r_1$ it suffices to know the value of $\re$ at that instant. A
lower limit for this value is obviously zero, so
solving the inequality (\ref{eq:fluxint}) for $r_1$ we get
\begin{equation} r_1< -\frac 32 \re+\left[\frac 94 \re^2 +3\frac{B_0}{\Be}(\rphio^2-\re^2)
   \right]^{1/2} < \left(\frac 94 \re^2 +3\frac{B_0}{\Be}\rphio^2
   \right)^{1/2}    \;.    \label{eq:wlimit} 
\end{equation}

\noindent Now, in the high-$\alphanu$ limit, the current sheet is very thin,
so that 
$r_e < \rphio$. This yields
\begin{equation} \frac{r_1}{\rphio} < \left(\frac 94 +3 
   \frac{B_0}{\Be}\right)^{1/2}\;.
   \label{eq:r1limit1} 
\end{equation}
Using (\ref{eq:wr1}) we obtain the following lower bound for $w$ in the limit
$\B0 \gg \Be$: 
\begin{equation} w > \frac{\nu_0}{\rphio}\frac 1{\sqrt 3} 
   \left(\frac{\Be}{B_0}\right)^{3/2} \label{eq:wlimit1} 
\end{equation}

We note without going into similar details that replacing the above straight
line with a logarithmic profile [stationary solution of the linear
diffusivity equation with the boundary condition $B(\re)=\Be$] would lead to
a transcendental equation instead of the second-order equation
(\ref{eq:fluxint}). Its solution yields a condition nearly identical to
(\ref{eq:wlimit1}) without the factor $\sqrt 3$. How good this lower bound
for $w$ is can be seen in Figures~\ref{fig2} and \ref{fig3}, where the
dashed line represents the right-hand-side of equation (\ref{eq:wlimit1}) 
(or its
inverse, in the case of Fig.~\ref{fig3}), without the $\sqrt{3}$ factor. 

A more realistic estimate of the value of $\re$ at the time when the erosional
solution sets in can be given taking into account
that the initial transient phase physically consists of the period
when the diffusive {\it gnawing} of the outer parts of the tube increases the
gradient further in. Thus, 
one may expect the duration of the transient phase to be 
$t_{{\mbox{\scri tr}}}\sim\rphio^2/\nu_0$, 
i.e.\ the diffusion timescale. With this, one has
approximately $\re(t=t_{\mbox{\scri tr}}) \sim \rphio (1- w\rphio/\nu_0)$. 
Inserting this condition into the inequality (\ref{eq:wlimit1}) 
(or, better still, into the equivalent 
inequality for the logarithmic profile discussed in the previous paragraph),
one obtains a relation between $w$ and $r_1$. This relation, coupled with
equation (\ref{eq:wr1}), can be solved to find that in the limit 
$B_0\gg \Be$
\begin{equation} w\ga  2^{-1/3}\frac{\Be}{B_0}\frac{\nu_0}{\rphio}\;.
   \label{eq:wlimit2}
\end{equation}
This new lower bound (represented as a solid line in Figs.~\ref{fig2} and
\ref{fig3}) turns out to give a very good approximation to the numerical
results for $w$. 
In fact, the value of the coefficient $2^{-1/3}$ in
equation~(\ref{eq:wlimit2}) shows only weak sensitivity to arbitrary 
order-of-unity factors in the expression of $t_{\mbox{\scri tr}}$ and in the
upper limit used in the integral constraint (typical changes are by $\pm 0.1$).
Equations (\ref{eq:wlimit2}) and (\ref{eq:wr1}) then clearly
also imply that $r_1$ and $\rphio$ are of the same order of magnitude.

Closing this subsection we should stress that the validity of the results derived here is not confined to
the particular choice of the $\nu (B)$ function given in equation~
(\ref{eq:nuexpr}), as this particular form was nowhere used. For more general
profiles, the parameter $\alphanu$ may simply be defined so that the
relation
\begin{equation} \frac{\Be}{\nu_0}\left(\dv{\nu}{B}\right)_{\Be}   
                 =-\alphanu /4 
\end{equation}
remains valid, i.e.\ $\alphanu$ simply parameterizes the steepness of the
diffusivity cutoff at $\Be$. It is easy to see that all the
above results remain valid in this more general case.

\subsection{Current sheet thickness} \label{sec:sheet} 
Both in the numerical results of Sect.~\ref{sec:numsol} and in the
propagating solutions of Sect.~\ref{sec:plpar} the thickness of the current
sheet rapidly decreases for increasing $\alphanu$. In this section we obtain
an analytical estimate for this dependence and show that, in fact, the
current sheet thickness has a power-law dependence on the internal field
strength, $\B0/\Be$, with exponent $-\alphanu$.  
We show this for the analytical solutions of Sect.~\ref{sec:plpar}. The
result is then immediately applicable to the numerical erosional solutions
with high $\alphanu$, given the excellent agreement between them. 

In the current sheet, the field distribution must go through an inflection
point. Define the thickness of the current sheet as the scalelength of $B(r)$
at that point. Using equation~(\ref{eq:propsoln}), we get 
\begin{equation} 
\Delta \equiv 
 -{\Bi}\,\left(\dv{B}u\right)^{-1}_i =  
-\frac{\Bi}{w} \, \left(\dv{\nu}B\right)_i^{} =
 \frac{\nui^2}{\nu_0} \frac{\alphanu}{w}
 \left(\frac{\Bi}{\Be}\right)^{\alphanu}\;, 
\end{equation}
where the index $i$ refers to values at the inflection point. 
Substituting here (\ref{eq:wlimit2}), 
\begin{equation} \frac{\Delta}{\rphi}\;\lesssim\;
   2^{1/3}\;\alphanu \left(\frac{\nui}{\nu_0}\right)^2 
\frac{B_0}{\Be} \left(\frac{\Bi}{\Be}\right)^{\alphanu} \;.  \label{eq:sheet}
\end{equation}

\noindent Next we show that in the high--$\alphanu$ limit the inflection
point occurs high up in the current sheet, i.e., $\Bi \rightarrow
B_0$. At the inflection point one has $\frac{d^2 u}{dB^2} = 0$; using the
general integral [eq.~(\ref{eq:intforminv})] of the plane--parallel
form--invariant propagating solution, this implies
\begin{equation} \frac 1\nui \dv{\nui}B=\frac{1}{\Bi-B_0} \;.
\end{equation}
Substituting here the expression (\ref{eq:nuexpr}) of $\nu(B)$ we find
\begin{equation} \alphanu\left( \frac{B_0}{\Bi}-1\right)
   =1+\left(\frac{\Bi}{\Be}\right)^{-\alphanu} \;,
\end{equation}
the high-$\alphanu$ limit of which is indeed
\begin{equation}
\Bi\rightarrow\frac{B_0}{1+\tfrac{1}{\alphanu}}\rightarrow B_0 \;.
\end{equation}
Comparing this with equation (\ref{eq:sheet}), and using the limit $\B0 >>
\Be$, we finally obtain:
\begin{equation} \frac{\Delta}{\rphi}\lesssim
   2^{1/3}\;\alphanu \left(\frac{B_0}{\Be}\right)^{1-\alphanu}   \;.
\label{eq:thickness}
\end{equation}
which was to be shown.

\section{Applications}\label{sec:applications}
The solar magnetic field appears in the form of magnetic flux tubes of
different sizes in the solar atmosphere. Through the studies of the emergence
of the active region fields as well as from general theoretical results of
magnetoconvection and MHD turbulence, one assumes that also the magnetic
field in the convection zone may be in the form of individual flux tubes or
bunches of them. For the application of the one-dimensional results of the
foregoing sections to the Sun, the magnetic flux tubes must be, even if only
approximately, cylindrically symmetric. In our case, a necessary condition
for this is that the lengthscale of variation of properties along the tube
axis be much larger than transversely to it (e.g., the tube radius must be
much smaller than all local stratification scaleheights). Under this
condition [typical, for instance, of the so--called thin-flux-tube ({\it tft})
approximation], equation~(\ref{eq:diff}) can be used to describe 
approximately the
diffusion of field through the surrounding turbulence. We stress that while
the  assumption of cylindrical symmetry is a common feature of our models and
of the {\it tft\/} formalism, yet here we break away from the {\it tft\/}
approximation by explicitly considering the field variation across the tube.
In the following we consider two cases of current interest, first one in
which the cylindrical shape of the tube is a good approximation, namely, the
magnetic tubes in the deep convection zone, and, second, the diffusion of the
field in sunspots. Sunspots can be simplified using axially symmetric models;
yet, they do not possess cylindrical symmetry, so this case can only
be seen as a simple approximation to a far more complex problem.

\subsection{Toroidal flux tubes at the bottom of the solar convective zone}
\label{sec:toroidal}
It is generally accepted that most of the magnetic flux responsible for solar
activity phenomena in general, and for the appearance of active regions in
particular, is stored close to the bottom of the solar convective zone
(\citeNP{Spiegel+Weiss:Nature}; see also \citeNP{FMI+:storage}). The
question arises whether these toroidal flux bundles are affected by the
phenomenon of turbulent erosion modelled above to any significant degree.
Using equation (\ref{eq:wlimit2}), the flux loss rate from a tube is
\begin{eqnarray}
  -\dot\Phi=2\pi \rt B_0 w &=2^{2/3}\pi\nu_0 \Be
   \rt/r_0 =2^{2/3}\pi\nu_0 \Be
   (\Phi/\Phi_0)^{1/2} 
\label{eq:decayrem} 
\end{eqnarray}
where we used the approximation $\rt\simeq (\Phi/\pi B_0)^{1/2}$, 
and we took into account
that $\rt$ and $\Phi$ are functions of time, while $w$ is determined
by the 
{\it initial \/} radius $r_0$ (or equivalently $\Phi_0$) in the case when the
tube is not moving compared to its surroundings. In a situation where a tube
is moving through the surrounding plasma (as in the case of {\it emerging \/}
flux loops) with a significant speed ($V>\nu_0/\rphi$), the external flow
relative to the tube will sweep away the weak outer field thereby
continuously ``reinitializing'' the decay; in that case, the {\it actual},
time-dependent value of $\rt$ should be used in the expression for $w$, and
consequently equation~(\ref{eq:decayrem}) must be replaced by:
\begin{equation}
    -\dot\Phi =2^{2/3}\pi\nu_0 \Be \qquad \mbox{(moving tube)}
      \label{eq:decaymoving}
\end{equation}

Consider first the implications of equation~(\ref{eq:decayrem}) for
magnetic tubes in the deep convection zone. The full turbulent
diffusivity there 
is of order $\nu_{00}\sim WH\sim 1000\,\mbox{km}^2/$s, ($W\simeq
60$\,m/s is the r.m.s.\ vertical velocity of turbulence), but as the scale of
the turbulent eddies is the pressure scale height $H\sim 5\cdot 10^4\,$km, 
i.e.\ much larger than the tube size, the diffusivity should be rescaled as
\begin{equation} \nu_0=\nu_{00}\left(\frac{\rphi}{H}\right)^{4/3}
   \sim W \left(\frac\Phi{\pi B_0}\right)^{2/3}H^{-1/3} .
   \label{eq:nurescale} 
\end{equation} 
Writing this into equation (\ref{eq:decayrem}) we find

\begin{equation} 
\dot\Phi= -W\,\Be\,\left(\frac{4 \pi \Phi^2}{H \B0^2}\right)^{1/3} 
            \left(\frac{\Phi\ }{\Phi_0}\right)^{1/2} \;.
\label{eq:decaynorem2}
\end{equation}

Numerical simulations of the emergence of buoyant magnetic flux loops through
the solar convective zone
\cite{FMI:classic,D'Silva+Choudhuri,Caligari+:movie,Fan+:loops2} show that in
order to get a satisfactory agreement with the observations of solar active
regions, the toroidal flux tubes on which the perturbations develop must have
magnetic flux densities of $B_0\sim 10^5\,$G, while the equipartition field
at the bottom of the convective zone is $\Be\sim 10^4\,$G. Those models do
not take into account any diffusivity, so, to match the observed fluxes in
active regions, the tubes must have magnetic fluxes of order $10^{22}$ Mx
also in the deep convection zone. With these values, using equation
(\ref{eq:decaynorem2}) we find that $-\Phi/\dot\Phi\sim 1$ month for the flux
loss timescale.  This is very much shorter than the length of the solar
activity cycle.  While the timescale increases with time owing to the factor
$(\Phi/\Phi_0)^{1/2}$ in equation (\ref{eq:decaynorem2}), the greater part of
the original flux will still be lost on essentially the initial timescale.

On the basis of such short lifetimes, it appears that flux tubes with the
above properties cannot be stored in the lower part of the convective zone
for times comparable to the solar cycle.  We note however that $\dot\Phi$,
and consequently the timescale, depend rather sensitively (as $W^2$, since
$\Be\propto W$) on the r.m.s.\ turbulent velocity. Thus, reducing $W$ by a
factor of $5$--$10$ could increase the decay timescale to several years,
comparable to the solar cycle. 

However, a careful analysis of the possible sources of error, taking into 
account the results of the relevant numerical simulations 
(\citeNP{Chan+Sofia:apj}, \citeNP{Lydon+:conv1}, \citeNP{Kim+:2}) 
shows that the $W$ values derived here are subject
to uncertainties not exceeding $\pm 30$--$40\,$\%. In particular, there seems
to be no obvious way to reduce $W$ by more than about 20\,\% at the base of
the convective zone. So the flux storage problems mentioned above appear to 
be unavoidable.

Thus, the calibration of local convection
theories to numerical experiments leaves now little room to change $W$ in the
unstably stratified part of the convective zone. On the other hand, our 
present ignorance
regarding the detailed structure of the overshooting layer below still allows
the possibility of storing the toroidal flux there.

Turning now to expression (\ref{eq:decaymoving}), we consider the loss of flux
in unstable magnetic tubes that rise across the convection zone to produce
active regions. To that end, we substitute for $\nu_0$ using equation~
(\ref{eq:nurescale}) to get:

\begin{equation} 
\dot\Phi= -W\, \Be\,\left(\frac{4 \pi \Phi^2}{H \B0^2}\right)^{1/3}  
\;.\label{eq:decaynorem4} 
\end{equation}

A first impression for the orders of magnitude predicted by this formula
can be obtained by substituting in it a few time profiles $z(t)$ and $\B0(t)$ 
typical of the {\it tft} numerical simulations of the rise of magnetic tubes
mentioned above, with $z$ the depth below the photosphere. Taking, for
instance, for $z(t)$ and $\B0(t)$ the position and field strength of the
summit of the rising loop in the simulations of \citeN{Caligari+:movie}, 
one can obtain $\dot\Phi$ from equation~(\ref{eq:decaynorem4}). 
To avoid the added complication associated with the sliding motion of the
tube's mass elements along the field lines, we use here time profiles 
corresponding to a non-rotating case ($\Omega=0$), for which the mass
element at the tube summit is fixed along time. On the other hand, 
one has to substitute for $W, H$ and $\Be$ using a stratification 
model and the following formul\ae: 
\begin{equation} W^3=\frac 1{1.37}\nablaad\frac{L_\odot}{4\pi R^2\rho}
   \label{eq:W} 
\end{equation}
\cite{Kim+:2}, where $\nablaad$ was simply taken to be 0.4 (a
good approximation except near the surface) and
\begin{equation}
\Be = 4\pi\rho\chi W^2 \;,
\label{eq:Be}
\end{equation}
with the numerical factor $\chi=1+2\cdot 0.61^2$ taken from
\citeN{Chan+Sofia:apj}. 

%\placefigure{figfluxloss}
\begin{figure}[htbp]
%\plottwo{fig1a.eps}{fig1b.eps}
%\plotfiddle{fig2.eps}{10 cm}{0}{100}{100}{0}{0}
\plotone{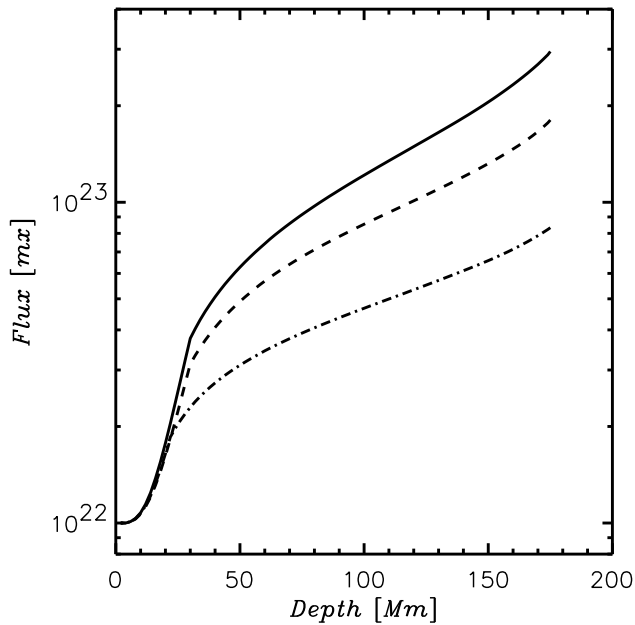}
\caption{
Flux loss from flux loops rising through the solar 
convective zone. Solid: $\Phi=10^{22}$ Mx, $B_{00}=1.2\cdot 10^5$ G. 
Dashed: $\Phi=10^{23}$ Mx, $B_{00}=1.2\cdot 10^5$ G.
Dash-dotted: $\Phi=10^{22}$ Mx, $B_{00}=2\cdot 10^5$ G. \label{figfluxloss}}
\end{figure}

Figure \ref{figfluxloss} shows the decrease of the magnetic flux during the
passage through the convective zone predicted by Eq.~(\ref{eq:decaynorem4})
for tubes of different sizes and field strengths. We first use the $z(t)$
profiles obtained from the thin-flux-tube numerical simulation of the rise of
tubes with $10^{22}$ Mx and $B_{00}=1.2\cdot 10^5$ G (solid curve) or
$B_{00}=2\cdot 10^5$ G (dash-dotted). We integrate Eq.~(\ref{eq:decaynorem4})
downward starting from the surface with that value of flux. We see that these
tubes must lose a substantial fraction of the their original magnetic flux
along the rise (in contrast to the simplifying assumption systematically made
in the thin flux tube simulations). The difference between the two curves is
due to the shorter timescale of the tube with higher initial field strength.
Anyway, both curves show flux values of order $10^{23}$ Mx in deep
levels. So, we repeat the calculation, this time taking $z(t)$ from a {\it
tft} simulation with $10^{23}$ Mx and $B_{00}=1.2\cdot 10^5$ G (dashed curve)
(but start the integration of Eq.~(\ref{eq:decaynorem4}) at the surface with
the same $\Phi$ value as for the other curves, for easiness of
comparison). Here too the tube suffers a substantial magnetic flux loss.
However, the calculation shows only a relatively small sensitivity to the
initial value of the magnetic flux in the emergence computation.  This is
because the magnetic flux of the tube only enters the {\it tft} calculation
via the drag force opposed by the surrounding medium to the advance of the
tube. In the range of fluxes of the figure, the radius of the cross section
is large enough for the drag force not to play a primary role in determining
the speed of rise - hence the time evolution profile of all quantities in the
tube is fairly independent of its magnetic flux. This is a fortunate
circumstance as it justifies {\it a posteriori} our (strictly speaking
inconsistent) approach of studying flux loss using models computed with the
assumption of a constant flux. It is perhaps also adequate to mention at this
point that the use of $z(t)$ and $\B0(t)$ profiles taken from numerical
simulations for a rotating case ($\Omega = \Omega_\odot$) yields results
basically coinciding with those shown in Figure~\ref{figfluxloss} within a
factor $2$ at most (yet, the interpretation of the results for the rotating
case has to be done with some care, as indicated above).

From all this we conclude that the toroidal flux tubes whose emergence leads
to the formation of solar active regions must lose a significant part of
their flux by turbulent erosion during their passage through the convective
zone. In fact, it would be interesting to include the formul\ae\ developed in
this section into a {\it tft} model, so that one could calculate the
time evolution of the magnetic flux in a more self-consistent way. Doing that
could result in a non-constancy of the magnetic flux of the tube as one moves
along its axis at any given instant of time. This would be no violation of
the solenoidality of the $\bf B$ field, but, rather, just reflect the fact
that the external sheets of a stretch of the tube may at some level detach
from its concentrated core and go into a weak-field phase that is no longer
counted as part of the tube in the {\it tft} model.

\subsection{Sunspots}   \label{sec:moat} 
The sharpness of the outer penumbral boundary of sunspots (i.e.\ the sudden 
change from filamentary penumbral fine structure to nearly normal granulation 
in less than a resolution element, \citeNP{Zwaan:ARAA,Gokhale+Zwaan}) 
is a well-known observational
fact that implies the presence of a current sheet around the spot
in the photospheric layers.  
Studies of sunspot structure \cite{Jahn} 
indicate that a significant part of the currents is also concentrated into a
current sheet in larger depths below the surface.  The calculations presented
above demonstrate how turbulent erosion of the surface of a magnetic flux
tube can lead to the spontaneous formation of a current sheet around the
tube. We thus propose that the sharp boundary of sunspots may be a natural
consequence of turbulent erosion. As observations indicate that pores and
sunspots tend to have well-defined boundaries from the 
beginning of their existence, this process of current sheet formation should
occur below the surface, before the eruption of the tube. 

With the model developed in this paper we can attempt to understand some
aspects of sunspot decay. That this process may be due to the eroding action
of external motions was in fact already proposed by
\citeN{Simon+Leighton:supgr.obs}.  However, one must keep in mind that
sunspots are very far from being slender flux tubes, so the application of
the present models may only be suggestive, and the resulting formulae very
approximative.

The constancy of $w$ implies a constant flux loss rate/tube surface area;
therefore, if one introduces an artificial sink term in the diffusion
equation outside the current sheet, thereby increasing the gradient and the
flux loss, the decay rate should increase. This was indeed found in a
numerical solution. A physical mechanism capable of
such flux removal from the neighbourhood of a flux tube could be a transverse
flow sweeping away the diffuse field (as with toroidal flux emerging through
the convective zone). Again, such processes could only increase the
effectiveness of turbulent erosion, thus the lifetime estimates given in
these applications should be regarded as rather conservative upper estimates.
On the other hand, such a removal of the outer diffuse flux may be regarded
as a ``renovation'' of the initial state, so that the end of the period of
effective renovation (i.e.\ the full emergence of the loop with
verticalization of the legs) may be regarded as the instant of time
corresponding to $t=0$ in our calculations. This may be approximately also
the time when the spot area attains its maximal value, so ``$r_0$'' may
represent the maximal radius of the spot.

Using equation~(\ref{eq:wlimit2}) the lifetime of a spot in the limit of strong
inhibition of turbulence is
\begin{equation} \rphio/w=2^{1/3}\frac{\rphio^2}{\nu_0}\frac{B_0}{\Be}
\end{equation}
Substituting here $B_0\sim 3000\,$ G, $\Be\sim 400\,$G,
$\nu_0\sim 1000\, \mbox{km}^2/$s (the granular value), one finds
\begin{equation} \rsp/w=(\rsp/10^4\,\mbox{km})^2\times 10^{\rm d}   \label{eq:spotlife}
\end{equation}
From this formula it is apparent that the total lifetime is proportional to
the initial area of the spot, thus returning the well known linear area-lifetime relation (\citeNP{Gnevishev}). 
The orders of magnitude are also consistent with
observations: in particular, the total lifetime of large, recurrent
spots of $\sim 5\cdot 10^4\,$km diameter is found to be about 2--3 months.

The area-time decay curves resulting from a current sheet moving inwards with
a constant velocity are necessarily parabolic, as can be seen in
Figure~\ref{fig1area} as well. This would run counter the common view that
the sunspot decay proceeds linearly in time.  There have been suggestions of
the parabolic nature of the sunspot decay laws on the basis of a statistical
study of large data samples (\citeNP{FMI+MVA}). The scatter intrinsic to this
kind of data, however, makes it difficult to distinguish a parabola with the
curvature predicted above from a linear decay law (\citeNP{VMP+:periph.decay}).
Nevertheless, in a recent analysis, \citeN{Petrovay+vDG:ASPE} found decisive 
observational evidence in favour of a parabolic decay law.

We note that in a parallel work \citeN{Rudiger+Kitchatinov} solved the \it 
2-dimensional\/ \rm nonlinear diffusion equation. 
As far as it can be judged from their figures, their 
solution falls in the diffusive regime, with a nearly linear decay law (in 
apparent contradiction to the observations, cf.\ \citeNP{Petrovay+vDG:ASPE}). 
As however only results from a single run with one particular choice 
of initial conditions are available, and the uppermost 1500 km of the convective zone (where $B/B_{\mbox{\scriptsize {e}}}$ would be highest) is not included in the 
model volume, it is presently not possible to judge the general validity 
of those findings.

An interesting property of the erosion models is that the {\it remaining
lifetime} of a sunspot with instantaneous radius $\rt$,
\begin{equation} \rt/w=(\rt/\rsp)
   (\rsp/10^4\,\mbox{km})^2\times 12^{\rm d} ,
\end{equation}
or, equivalently, the area decay rate
\begin{equation} 2\pi \rt w=\frac{\rt}{\rsp}\pi
   \frac{(10^4\,\mbox{km})^2}{6^{\mbox{\scri d}}} 
\end{equation}
does not only depend on $\rt$ but also on $\rsp$, i.e.\ on the {\it initial}
radius of the spot. 
%(or equivalently, on its total flux, including the part
%which has already diffused out of the spot). 
This is caused by the fact that the value of $w$ is fixed at an early stage
of the decay, when the propagating solution is first adopted, and  it
does not change much thereafter.  This implies that of two observed decaying
sunspots 
with identical radii, the younger one (i.e.\ the one with a smaller maximal
radius) should show a faster decay rate. One may speculate that this
dependence on the original size may possibly be a contributing factor to the
large intrinsic scatter in the decay rates of sunspots of a given size.

\begin{figure}[htbp]
%\plottwo{fig1a.eps}{fig1b.eps}
%\plotfiddle{fig2.eps}{10 cm}{0}{100}{100}{0}{0}
\plotone{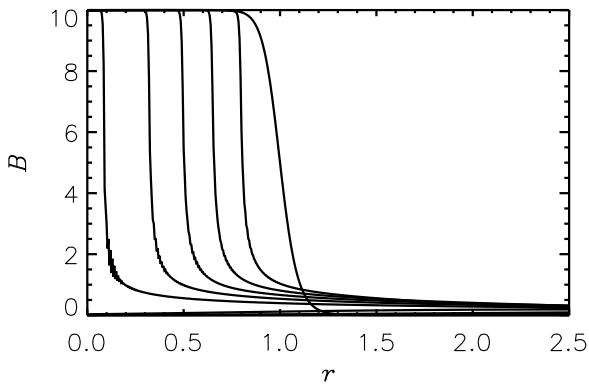}
\caption{
Same as Fig.\ 1$c$ but including a radial outflow as in eq.~(36), %(\ref{eq:moatflow}
with $v_0=1$, $\alpha_v=5$, $\Delta t_0=0.5$.
\label{fig1moat}}
\end{figure}

From observations it is known that sunspots are surrounded by a radial
outflow, called the moat. It is then interesting to investigate the effects of
such an outflow on the solutions. The actual structure of the moat flow is
not totally clear: since the field lines are inclined, the flow velocity
could have a non-negligible component parallel to the field lines. 
However, only the component normal to the field may have a direct effect on
its decay. We then use a more general form of the diffusion equation including
an advection term due to the radial outflow of matter, and, 
% more general form of Eqn.~(\ref{eq:diff}) introducing an advection
%term due to a flow in the radial direction, and, 
within the one-dimensionality of the model, disregard any effects due to any
other component of the flow. The resulting equation is: 
\begin{equation}
\pdv{}t\left(rB\right)=\pdv{}r\left[r\;\nu(B)\;\pdv{B}r + v_r\,r\,B\right] \;,
\label{eq:advection}  \end{equation}
\noindent which substitutes equation~(\ref{eq:diff}). Observations suggest
that outside the penumbra the outflow has a 
velocity of $\sim 0.5$--$1$\,km/s, nearly independently of the radius.  
\cite{Pardon+,Brickhouse+LaBonte:moat.obs}.  On the other hand, the radial
flow must have a sharp cutoff at the boundary of the strong field region. 
The function $v_r$ may then be represented as 
\begin{equation} v_r=\frac{2v_0}{1+(B/\Be)^{\alpha_v}} \label{eq:moatflow}
\end{equation}
where $v_0$ is the outflow velocity at $\re$. Equation (\ref{eq:wlimit2}) can be
readily generalized to yield
\begin{equation} w\ga  2^{-1/3}
  \frac{\Be}{B_0}\left(\frac{\nu_0}{\rphi}+v_0\right)\;.\label{eq:wwithflow}
\end{equation}
We have obtained numerical solutions for this problem: the resulting field
profile can be seen in Figure \ref{fig1moat}, which presents the case
for $v_0=1$, $\alpha_v =5$, $\alphanu =3$. These solutions confirm the
accuracy of the estimate (\ref{eq:wwithflow}). 
Now the dimensionless value of $v_0$ will be 0.5\,km/s$\times \rsp$/$\nu_0$,
i.e. crudely in the range $2<v_0 \rsp/\nu_0<15$. Consequently, a moat 
flow perpendicular to the field lines would imply an order-of-magnitude
increase in the decay rate, in conflict with the observations. This leads us
to conclude that, to the extent that the present models can teach us
something about the process of sunspot decay, the moat flow should be mostly
parallel to the field lines. This conclusion also agrees with the
observations of \citeN{Skumanich+} that the {\it apparent} flux transport
rate of the moat (in the form of moving magnetic features, MMFs) greatly
exceeds the actual flux loss from the spot.

%\placefigure{fig1moat}

\section{Conclusion}
An investigation by numerical and analytical means of the solution of 
the nonlinear axisymmetric diffusion 
equation (\ref{eq:diff}) with $\nu (B)$ functions that satisfy the minimum set
of physical requirements listed in \S\ref{sec:formulation} yielded the
following main results. For sufficiently strong nonlinearity (i.e., for
sufficiently strong  
reduction of $\nu$ inside the tube) a current sheet is spontaneously formed 
around the tube within one diffusion timescale $r_0^2/\nu_0$ ($r_0$ is the 
initial radius of the tube, and $\nu_0$ is the kinematical value of the 
diffusivity).  The field profile in and inside the current sheet is well 
approximated by the analytical form-invariant propagating solution of the
plane-parallel equivalent of equation (\ref{eq:diff}). This sheet propagates
inwards with a velocity 
\begin{equation} w\sim  2^{-1/3}\frac{\Be}{B_0}\frac{\nu_0}{\rphi}
\end{equation}
where $B_0$ is the field strength just inside the current sheet, $\Be$ is the
equipartition field strength. Accordingly, the lifetime of a tube  
with constant internal flux density is increased by a factor $B_0/\Be$, {\it
independently of the value of the diffusivity inside the tube}.

On the basis of these results we performed approximate calculations of the
magnetic flux loss from toroidal flux tubes lying at the bottom of the solar
convective zone, and rising through the zone to the surface, respectively. It
was found that the timescale of flux loss is $\sim 1$ month, comparable to
the rise time, and very much shorter than the solar cycle. Consequently,
toroidal flux bundles cannot be stored inside the convective zone proper for
extended periods of time and the layer of flux storage must have turbulent
diffusivities (and therefore turbulent velocities) significantly lower than
the convectively unstable layer. Flux loops rising through the convective
zone lose a significant fraction of their magnetic flux during the rise. 

While sunspots are far from being thin flux tubes, an application of our
models to them still yields decay times comparable to those observed;
besides, the linear area-lifetime relation is also returned. The inclusion of
a moat flow with the observed velocities and perpendicular to the field lines
however reduces the decay times to values incompatible with observations;
thus, the moat flow may actually rather be more or less parallel to the field
lines. Furthermore, turbulent erosion may explain the origin of the current
sheet at the boundary of a sunspot. A curious feature of our model is that
the decay rate does not only depend on the actual size of a spot but also on
its original (i.e.\ maximal) size. This is the signature of a parabolic decay
law only slightly departing from linear. Observations seem to confirm our
prediction of a parabolic decay law (\citeNP{Petrovay+vDG:ASPE}). 
On the other hand, from the solutions obtained in this paper, we can estimate
the Joule dissipation associated with the current sheet around a flux tube.
Setting in the values corresponding to a sunspot, we obtain 
a photospheric power source that is 4 orders of magnitude lower than
the solar radiative output on the corresponding area (an annulus of thickness
$1"$ around the spot), far too weak to be observed. 

On the theoretical front, possible extensions of this work include a
generalization by allowing the variation of physical quantities in the
direction parallel to the tube, as well as tests of these results in
three-dimensional MHD simulations.

\acknowledgements

The authors are grateful to Dr Manfred Sch\"ussler for useful comments on
this manuscript. This work was financed in part by the DGES project
no.~95-0028 and by the OTKA under grant no.~F012817.

%\input erosion.bbl

%\bibliography{apjmnemonic,kris0}
%\bibliographystyle{apj}

%\clearpage

\end{document}